\documentclass[twocolumn,showpacs,showkeys,aps,amssymb,floatfix,prd,amsmath,preprintnumbers,10pt]{revtex4-2} 
\usepackage{graphicx}  
\usepackage{dcolumn}   
\usepackage{bm,color}
\usepackage{lineno}
\usepackage{hyperref}
\begin{document}

\def \reals{{\mathbb R}}
\def\be{\begin{equation}}
\def\ee{\end{equation}}
\def\bea{\begin{eqnarray}}
\def\eea{\end{eqnarray}}
\def\nn{\nonumber}
\def\th{\theta}
\def\ph{\phi}
\def\lt{\left}
\def\rt{\right}
\def\ed{\end{document}}
\def\degree{\mathop{\rm {{}^\circ}}}
\input epsf.tex

\title{Compactness of supermassive dark objects at galactic centers}\thanks{This paper is  dedicated to the memory of my {\em guru}  Professor P.~C.~Vaidya. (In Sanskrit, {\em gu} means darkness and {\em ru} means one who dispels.)}

\author{K. S. Virbhadra}
         \email[Email address: ]{shwetket@yahoo.com}
         \affiliation{Mathematics Department, Drexel University, 33rd and Market Streets, Philadelphia, PA 19104, USA}
         \homepage[Visit: ] {https://www.math.drexel.edu/~ksv29/}

\begin{abstract}
We define compactness of a gravitational lens  as the scaled closest distance of approach (i.e., $r_0/M$) of the null geodesic giving rise to an image. We model forty supermassive dark objects as Schwarzschild lenses and compute compactness of lenses (determined by the formation of the first order relativistic image). We then obtain a novel formula for the compactness of a lens as a  function of mass to the distance ratio ($M/D_d$) and the ratio of lens-source to the observer-source distances ($D_{ds}/D_s$). This formula yields a very important result: Just an observation of a relativistic image  would give an incredibly accurate  upper bound to the physical  compactness  (the ratio of the radius to mass) of the lens without having any knowledge of mass of the lens, angular source position, and observer-source and lens-source distances. Similarly, we show that the observation of the second order relativistic image would give a lower value of upper bound to the physical compactness. These results, though  obtained for supermassive dark objects at galactic centers, are valid for any object compact enough to give rise to relativistic images.
\end{abstract}



\maketitle

\section{\label{sec:Intro}Introduction}

In the year 2000, we obtained a new gravitational lens equation that allows arbitrary large light deflection angles and studied Schwarzschild lensing. Apart from the usual primary (also called direct) and secondary images, we got a theoretically  infinite sequence of relativistic images on both sides of  the optic axis \cite{VE00}. The relativistic images are formed due to light deflection in a very strong gravitational field in the vicinity of the photon sphere and  are incredibly demagnified unless the lens components are highly aligned and the  lens-source distance is small. Due to such a big observational difficulty, we were sheepishly optimistic that relativistic images would be observed in the near future and considered that a Herculean task for astronomers to accomplish. Historically, researchers have studied the law of nature from the small scale to the large scale due to their  curiosity rather than any hope for
immediate practical applications and this is how theoretical physics has progressed. Therefore, a large  number of researchers  pursued  research in gravitational lensing in the strong gravitational fields of black holes \cite{V09,Pan23,Pan22,Tsu23,Tsu22,Wal23,Ash23,Kob22,IshEtal16,IshEtal17,Ono17,Tak20,Kum22,Ata21,Gao21,Tsup21,Kli19,Zha17,Ish17,You15,Vir22, Adl22,Vir24},  naked singularities \cite{VNC98,VK08,VE02,Tsu21,Gyuetal19,OSZ15,Sahetal12}, and  wormholes and other exotic  and esoteric objects \cite{Izmetal19, SR20,SI15,GX21,ZJC17,SFM06}. (See also references therein.)

The concept of a photon sphere is well-known in the literature and its existence or non-existence in  spacetime plays a very important role in theoretical as well as observational astrophysics, especially in the context of gravitational lensing. In view of this,  we \cite{VE00} defined a photon sphere that has  a direct relation to gravitational lensing: 
A photon sphere in  static spherically symmetric spacetime is a timelike hypersurface $\{r=r_o\}$  in the spacetime if the Einstein bending angle of a light ray with the closest distance of approach   $r_o$   becomes unboundedly large. This definition yields  $r_o = 3M$  for the Schwarzschild metric, where the parameter  $M$ is the ADM  mass (also called the Schwarzschild mass.)
The null geodesic equation for the maximally extended Schwarzschild metric demonstrates that any future endless null geodesics at any point with  $r:2M<r<3M$ and initially directed inwards continue inwards and fall into the black hole. However, any future endless null geodesics starting at some point $r>3M$ and initially directed outwards will escape to infinity and can be captured in a telescope.
 Most  astrophysical objects cannot be expressed by static and spherically symmetric spacetimes. Therefore, later, we generalized the concept of a photon sphere to a {\em photon surface} and gave a rigorous geometrical definition (see \cite{CVE01} for details.) Any null geodesic initially tangent at a photon surface and starting at the photon surface will remain in the photon surface.  As expected, we found that our two different definitions gave the same result for a general static spherically symmetric metric \cite{CVE01,VE02}. We also proved a few important theorems that are likely to have immense implications for astrophysics. We   also proved that subject to an energy condition, a black hole described by a  static spherically symmetric spacetime must be surrounded by a photon sphere. However, no generalization of this theorem  for black holes described by more general spacetimes is  known.   The studies of photon surfaces, particularly photon spheres,  were further progressed by many researchers (see \cite{KIN21,CG21,YL15,GW16,MCS19,Wei20,KG20,Jah19} and references therein.)

Theoretical research on black hole lensing continued despite the very poor possibility of observation of relativistic images that are formed due to looping (deflection angles $\hat\alpha > 3 \pi/2$) of null geodesics around the photon sphere. We (the present author), in 2009, carried out a comprehensive study of Schwarzschild lensing and obtained exciting results not just related to the relativistic images, but also non-intuitive conceptually important results for the primary-secondary images, which were thought to be completely understood almost since the outset the gravitational lensing studies in the last century  \cite{V09}.
Though we do not wish to digress from the main topic of this paper, it is worth  mentioning here an  important result for the reason it will be clear in the next paragraph. The  distances of astrophysical objects play big obstacles in accurate  determination of their masses.  However, 
we obtained an incredibly accurate formula for the  mass of any Schwarzschild lens in terms of differences of time days  between relativistic images of the 1st and 2nd orders on the same side of the optic axis (see Eq. $19$ in \cite{V09}). The formula is extremely  insensitive to  the observer-lens and lens-source distances as well as the angular source position. To our knowledge, this is the most accurate  known formula for  obtaining masses of Schwarzschild lenses, which are  compact enough to produce relativistic images. 

Despite  huge publications on relativistic images, these remained unobserved. To this end, in April 2017, the EHT (Event Horizon Telescope) international collaboration mapped the supermassive compact core ($M87^*$ hereafter) of the giant elliptical galaxy $M87$ in the Virgo cluster. On April 10, 2019, they presented the  unprecedentedly resolved  image of the $M87^*$ with a big fanfare through six simultaneous press conferences as well as the publications of a series of six papers \cite{EHT6}. Thus, these paradigm shifting observations confirmed expected silhouette of the compact dark object  and light rings  in the vicinity of the photon sphere; the latter seem to be due to secondary as well as relativistic and mirror images. The present EHT is not capable of resolving  relativistic images from the secondary and mirror images and neither relativistic images themselves of different orders. However, we are hopeful that  the {\em ngEHT} \cite{ngEHT} will be able to  observe resolved relativistic images.  
This has  elevated the research enthusiasm of theoretical astrophysicists to pursue their studies of  black hole lensing. The existence of a photon sphere is a sufficient (but not necessary) condition for the formation of relativistic images \cite{V09}. Therefore, the  observation of relativistic images would very strongly support that the $M87^*$ is a {\it photohole} (a compact object covered within a photon surface or a photon ring), but it doesn't confirm it. (The term {\it photohole} is coined by the present author, and it appears for the first time here.)  And, therefore, we are far away from   iron-clad evidence for black holes. The observation of relativistic images however supports general relativity in a very strong gravitational field. Closer we can observe to the event horizon better the observation would  support the black hole interpretation. At this stage it is of interest to probe whether merely the observation of relativistic images gives a good   upper bound to the physical compactness (the ratio of radius to mass) of the massive objects that does not depend on the values of any lensing parameters  (mass of the lens, angular source position, and the lens-source and  observer-source distances). The aim of this paper is to find such an upper bound. In this paper, we show that just the observation of a relativistic image gives an extremely accurate value of the upper bound to the compactness of the massive compact object. This is similar to the  lens mass formula where distances play incredibly insignificant role. 

 The presentation  of this paper is arranged as follows. In Sect II, we review the lens equation and some other  equations that are used in computations in this paper. In Sect. III, we perform computations and present results. In the last section IV, we discuss and conclude the results. Throughout this paper, we use geometrized units ($G=1$ and $c =1$)  so that the mass $M$ of the lens is equivalent to  $GM/ c^2$.

\section{\label{sec:LE}Lens Equation and Schwarzschild  Lens}

A gravitational lens equation that allows arbitrary large deflection angles in a strong gravitational field is given by \cite{VE00}
\begin{equation}
  \tan\beta =  \tan\theta -  {\cal D} \lt[\tan\theta + \tan\lt(\hat{\alpha} - \theta\rt)\rt] 
  \label{GravLensEqn}
\end{equation}
with
\begin{equation}
  {\cal D} =
    \frac{D_{ds}}{D_s}   \text{,}
   \label{calD}
\end{equation}
where $\beta$ and $\theta$, respectively, stand for angular positions of an unlensed source and an image both measured from the optic axis (the line joining the center of mass of the lens to the observer). $\hat{\alpha}$ is the Einstein deflection angle of  the null geodesic.  $D_{ds}$ and $D_s$ denote, respectively, lens to source and observer to source angular diameter distances. The dimensionless parameter ${\cal D}$, the ratio  of the lens-source to observer-source distances, has values in the open interval $\lt(0,1\rt)$; however, in order for the lens equation to yield results with high accuracy, its value should not be taken too small. The impact parameter
\begin{equation}
J = D_d \ \sin\theta  \text{,} 
  \label{J}
\end{equation}
where $D_d$ stands for the observer to lens angular diameter distance.
We will model the supermassive compact  objects at galactic centers as Schwwrzschild lenses. The exterior gravitation field of Schwarzschild lens is given by the line element:
\begin{eqnarray}
ds^2&=&\lt(1-\frac{2M}{r}\rt)dt^2- \lt(1-\frac{2M}{r}\rt)^{-1} dr^2 \nonumber \\
      &-&r^2\lt(d\vartheta^2+\sin^2 \vartheta d\phi^2\rt) \text{.}
   \label{SchSpaceTime}
\end{eqnarray}
The real constant parameter $M$ is the ADM mass. Weinberg \cite{Wei72} obtained the bending angle of a light ray with the closest distance of approach $r_o$ which is expressed as 
\begin{equation}
    \hat{\alpha}\lt(r_o\rt) = 2 \  {\int_{r_o}}^{\infty}
                         \frac{dr}{r \  \sqrt{\lt(\frac{r}{r_o}\rt)^2  \lt(1-\frac{2M}{r_o}\rt)
                        -\lt(1-\frac{2M}{r}\rt)}   } - \pi \text{.}
   \label{AlphaHatR0}
\end{equation}
The impact parameter  of the light ray with the closest distance of approach $r_o$ is 
\begin{equation}
J\lt(r_o\rt) = r_o \lt(1-\frac{2M}{r_o}\rt)^{-1/2} .
       \label{ImpParaR0}
\end{equation}
We define a {\em compactness parameter} as follows:
\be
{\cal C} = \frac{r_o}{M} \text{.}
\ee
A light ray from a distant source reaching the telescope with the closest distance of approach $r_o$  to the lens confirms that the radius of the compact object is less than $r_o$.  
Ironically, for a lower value of $r_o$, the compactness parameter ${\cal C}$ is lower though the lens is more compact. Thus, for a more compact lens, the compactness parameter is smaller.
(Defining  the compactness parameter as $M/r_o$ does not give  physically appealing  real numbers  in order to  conveniently compare with the radius $3M$ of the  photon sphere.) 
We denote the compactness of the lens due to observations of the zeroth, first, and second orders ring images, respectively, by  ${\cal C}_o$,  ${\cal C}_1$ and ${\cal C}_2$. (The primary-secondary images are of the zeroth order. The outermost  relativistic images  are of order $1$ and  those next to the outermost are of order 2.)

\section{\label{sec:Comp}Computations and Results}


\begin{figure*}[tbh]
\centerline{ \epsfxsize 17.2cm
   \epsfbox{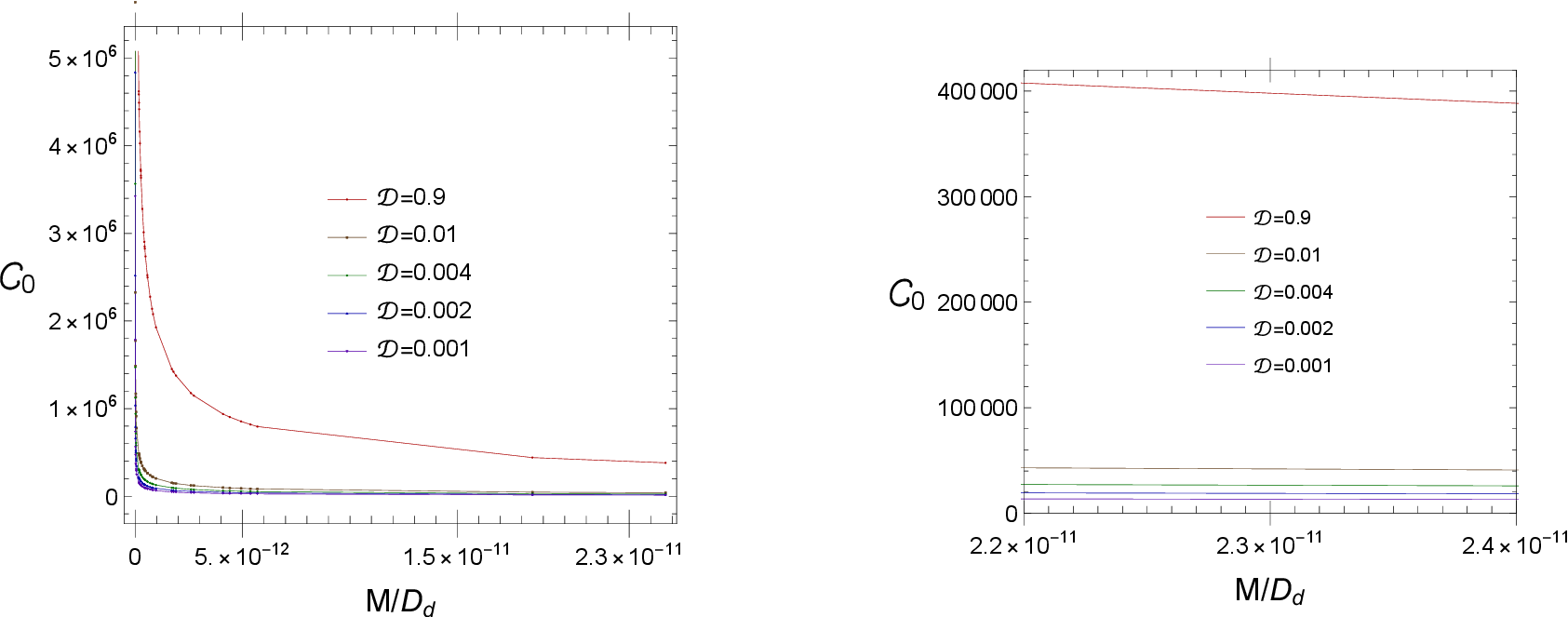}}
 \caption[]{
{\em Left}:   The compactness ${\cal C}_0$ of the lens (determined by the observation of the Einstein ring) vs $M/D_d$ (the ratio of the mass of the lens to the lens-observer distance) is plotted for different values of the dimensionless parameter
        ${\cal D}$. SMDOs at the centers of 40 galaxies are modeled as the Schwarzschild lenses. 
   {\em Right}: The magnified graphs of the figure on left  around the large value of  $M/D_d$ are shown. The curves for different values of  ${\cal D}$  are in the decreasing order of ${\cal D}$ (the highest ${\cal D}$-value curve is  on the top.)
   }
\label{fig1}
\end{figure*}

\begin{figure*}[tbh]

\centerline{ \epsfxsize 17.2cm
   \epsfbox{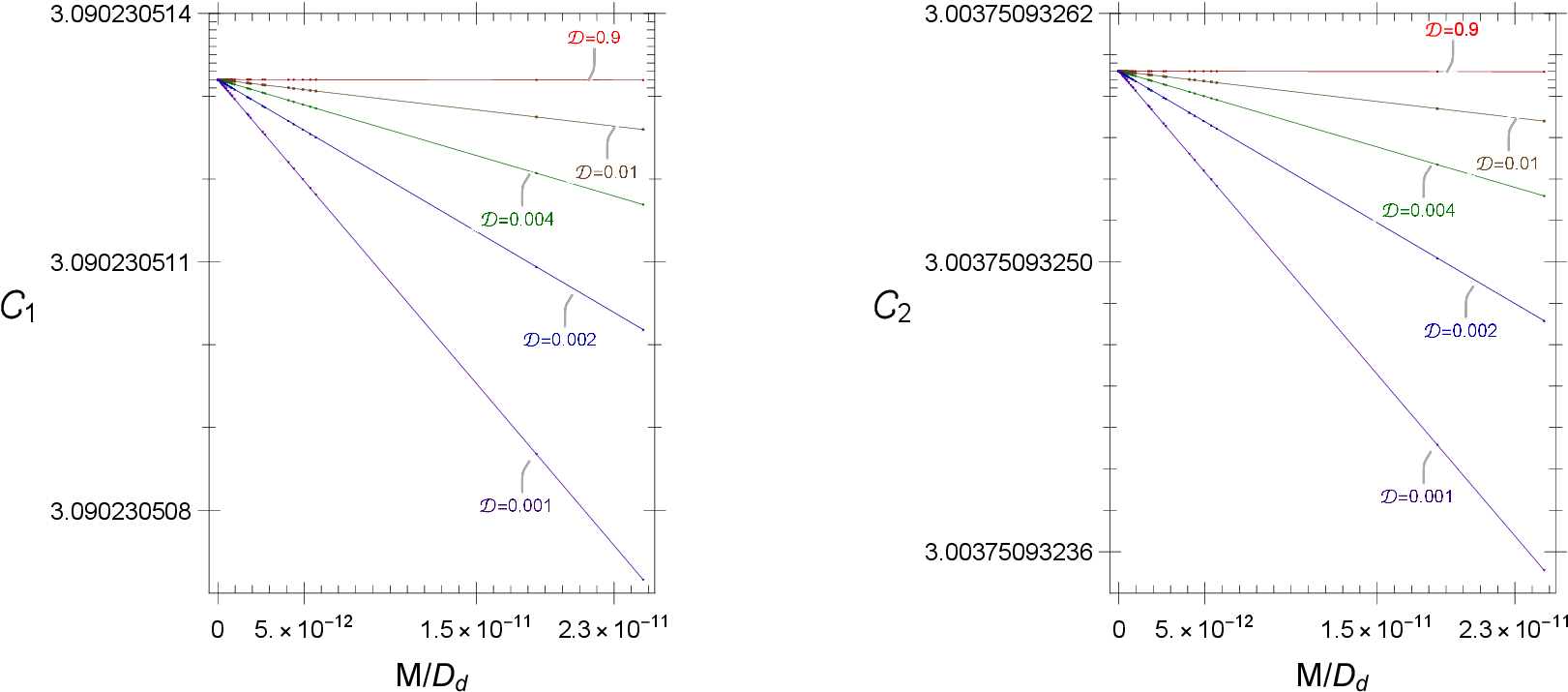}}
 \caption[ ]{ 
The compactness ${\cal C}_1$  and ${\cal C}_2$ of the lens (determined by the observations of, respectively,  the  first and second-order relativistic rings) vs $M/D_d$ (the ratio of the mass of the lens to the lens-observer distance) are plotted for different values of the dimensionless parameter
        ${\cal D}$. SMDOs at the centers of 40 galaxies are modeled as the Schwarzschild lenses.
}
\label{fig2}
\end{figure*}

\begin{figure*}[tbh]
\centerline{ \epsfxsize 14cm
   \epsfbox{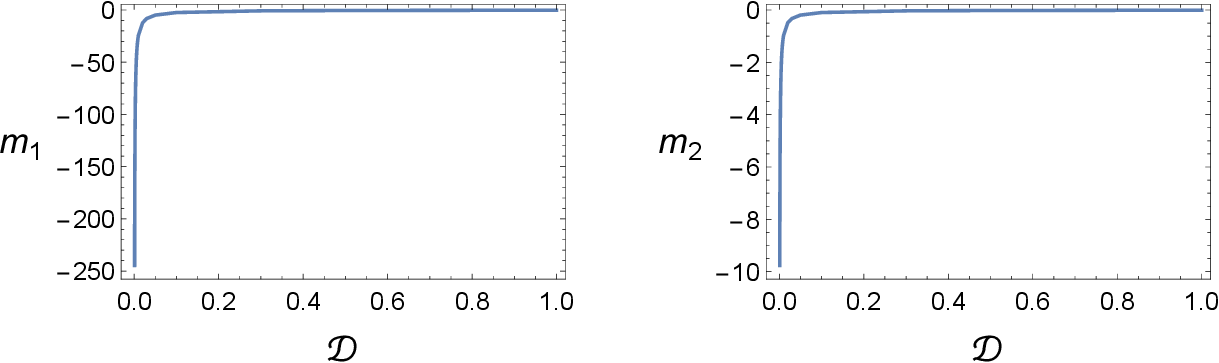}}
 \caption[ ]{
The slopes of compactness vs $M/D_d$ (the ratio of the mass of the lens to the lens-observer distance) graphs  are plotted against ${\cal D}$ (the ratio of the lens-source to the observer-source distances). The slopes $m_1$ and $m_2$ stand, respectively, for the cases of the first and second-order relativistic rings. 40 values of   ${\cal D}$  on the interval of $(0,1)$ are taken to plot these graphs.
}
\label{fig3}
\end{figure*}
We model SMDOs  (supermassive dark objects) at centers of $40$ galaxies (including the Milky Way and  $M87$) as Schwarzschild lenses and study point-source gravitational lensing by them. For the list of galaxies (with the masses and distances of the SMDOs at their centers), see Table IV in \cite{V09} and references in the table caption. Though the final results in this paper do not depend on the masses and distances of SMDOs, we use recently obtained values of masses and distances of  $M87^*$ \cite{EHT6} and $SgrA^*$ \cite{Abuter20}. We use {\it Mathematica} \cite{Math12} for entire computations in this paper. We begin with the ratio of lens-source to the observer-source distances ${\cal D} = 0.001$ and solve gravitational lens equation $(\ref{GravLensEqn})$ for 40 different lenses (i.e., different values of $M/D_d$ of SMDOs) and obtain scaled closest distances of approach $r_0/M$ of light rays for the formation of Einstein rings for each case. (We defined $r_0/M$ as the compactness of the lens.) We repeat these computations with  ${\cal D} =  0.002, 0.004, 0.01$, and, $0.9$. In  Fig.$\ref{fig1}$, we plot the compactness ${\cal C}_0$ vs $M/D_d$ for all these values of ${\cal D}$. For a given value of ${\cal D}$, the compactness ${\cal C}_0$ decreases with the increase in the value of $M/D_d$ and for a fixed value of $M/D_d$, the compactness ${\cal C}_0$ increases with the increase in the value of ${\cal D}$. As Einstein rings are formed due to light deflections in a weak gravitational field (far away from the lens), these investigations yield  very high values for compactness of  lenses. It is clear from  Fig.$\ref{fig1}$ that even when the source is relatively closer to the lens (i.e., ${\cal D} = 0.001$), the observation of an Einstein ring gives  a  large value of  the compactness  of the lens. Therefore, observations of Einstein rings are not a good way to determine an upper bound to the  physical compactness of SMDOs.
We now solve the lens equation to obtain positions of the first order relativistic rings and obtain compactness ${\cal C}_1$  of 40 MDOs for ${\cal D}= 0.001, 0.002, 0.004, 0.001$ and $0.9$.  In Fig.$\ref{fig2}$ (left), we plot  ${\cal C}_1$  vs $M/D_d$ for these values of ${\cal D}$. It is very fascinating that the family of  ${\cal C}_1$  vs $M/D_d$  graphs appear to be straight lines. We repeat this computation for the second-order relativistic images and plot ${\cal C}_2$  vs $M/D_d$ for different values of ${\cal D}$ and again get another family of curves that appear to be straight lines.  The curves do not intersect and show that their slopes are always negative and tend to  be $0$ as ${\cal D} \rightarrow 1$. Encouraged by these results, in order to have excellent curve fittings,  we carry out  computations for additional 45 values of ${\cal D}$ on the interval $(0,1)$, but not too close to endpoints of this interval. In Fig. 3, we plot the slopes of these two families of straight lines. Using {\it Mathematica} \cite{Math12}, we do curve fitting for both cases and obtain excellent results with  extremely small values for the diagnostic $1-RSquarred$ as well as very large negative values for the diagnostics $AICc$ (Akaike Information Criterion, where the subscript c stands for  a correction for small sample sizes) and $BIC$ (Bayesian Information Criterion). The entire computations are performed with very high accuracy.  We maintained $1-RSquarred < 10^{-30}$  and  AICc as well as  BIC both $<-2160$.  We obtain compactness ${\cal C}_1$ and   ${\cal C}_2$, which are determined, respectively, by observations of the first and second-order relativistic rings. These are given by
\bea
{\cal C}_1 &=&  3.0902305132011 + m_1 \  \frac{M}{D_d} \text{ and}  \nonumber\\
{\cal C}_2 &=&  3.0037509325921 +  m_2 \ \frac{M}{D_d} \text{,}
\label{C1C2}
\eea
where slopes $m_1$ and $m_2$ are, respectively, given by 
\begin{widetext}
\bea
m_1 &=&  \frac{7.4290887206379\times 10^{-12}}{{\cal D}^2} - \frac{0.2445976800315}{{\cal D}} \  \text{and}  \nonumber\\ 
m_2 &=&  \frac{2.7247061578677\times 10^{-13}}{{\cal D}^2} - \frac{0.0097615517187}{{\cal D}}  \text{.}
\label{m1m2}
\eea
\end{widetext}
(The reasons for putting a large number of digits are given  soon.)
An accurate determination of the mass of the lens as well as lens-source and observer-source distances has been always problematic. However, if we have an upper and lower bound to these, we can determine
excellent bounds (upper as well as lower) on the compactness of SMDOs. However, just by observation of relativistic rings (without any knowledge  of mass and distance), the above formulas give an upper bound (superscript {\em ub} used in the following) to the compactness of the SMDO. Higher the order of the observed relativistic image, we get lower and hence better value for the upper bound to the compactness. The equations $(\ref{C1C2})$ yield magnificent upper bounds on physical compactness of SMDOs by observations of relativistic rings of the first and second orders:
\bea
{\cal C}_1^{ub} &=&  3.0902305132011    \text{ and} \nonumber\\ 
{\cal C}_2^{ub} &=&  3.0037509325921  \text{.}
\label{ab}
\eea
Despite the fact that we found  highly accurate results for free (in the sense that we do not need to know masses and distances at all and no measurements are required)  the presentation of too many digits might appear ridiculous. Then, why did we do that? The following are reasons: (i)  Without that, the compactness vs $M/D_d$  plots will not be resolved and we will not be able to see the effects of the parameter ${{\cal D}}$. (ii) Despite the fact that general relativity is a non-linear theory, we obtained a surprisingly linear relation between the compactness and $M/D_d$ for a fixed value of ${{\cal D}}$. The presentation of a large number of digits supports linearity.   (iii) What is unmeasurable today could be measurable tomorrow as there is no upper bound to the success of future astronomy. 

Though we carried out computations for relativistic rings (i.e., with the angular source position $\beta = 0$), our results for bounds on the compactness are valid even if we observe a pair of relativistic images  on opposite sides of the optic axis or just a relativistic image of negative parity (i.e., on the secondary image side). Similarly, results obtained  here for the SMDOs  are applicable to compact objects of any mass giving rise to relativistic images.

\section{\label{sec:ConcludingRemarks}Concluding Remarks}

We modeled forty supermassive dark objects at galactic centers as Schwarzschild lenses and obtained the scaled closed distances of approach ($r_0/M$) of light rays producing rings of the zeroth-order (Einstein ring) for a few  values of $\cal{D}$ (the ratio of the lens-source to the observer-source angular diameter distances.) We defined $r_0/M$ as the compactness ${\cal{C}}$ of the lens. We first plotted the compactness ${\cal{C}}_0$ for the zeroth-order images (Einstein rings) and it gave a large value of compactness of the lens, as the zeroth-order rings are formed due to light deflection in a weak gravitational field that is far away from the lens. We repeated these computations for the first and the second order relativistic rings and plotted ${\cal{C}}_1$  vs $M/D_d$ and ${\cal{C}}_2$ vs $M/D_d$ for a few values of ${\cal{D}}$. We thus obtained a family of curves for each case, which appeared linear. 
The data showed uncannily linearity of these curves.
The compactness increases with the decrease in $M/D_d$ and increase in the value of $\cal{D}$. The family of curves do not intersect. The slopes of these straight lines are negative, increase with the increase in the value of $\cal{D}$, and approaches $0$ as ${\cal{D}} \rightarrow 1$, which is the right endpoint of the interval $(0,1)$. Inspired by the linearity of the families of curves, we repeated the computations for a total fifty values of ${\cal{D}}$ in the interval $(0,1)$ and then carried out curve fitting. We got formulas for curves for both the first and second order relativistic rings.
The observations of the first and the second order relativistic rings, respectively, give the compactness of the lens ${\cal{C}}_1$ and ${\cal{C}}_2$ which are functions of $M/D_d$ and ${\cal{D}}$. These expressions clearly give upper bounds,  $C_1^{ub}$ and $C_2^{ub}$ to the compactness $C_1$ and $C_2$, respectively. The results for upper bounds are valid  if  either the relativistic ring or the relativistic image on the secondary image side is observed. The upper bounds to the  physical compactness   of the lens do not require any knowledge of $M/D_d$, ${\cal{D}}$, and $\beta$.
However, if there is some knowledge available for $M/D_d$ and ${\cal{D}}$ (i.e., we have some bounds on these from other observations), we can easily compute better upper  and also lower bounds to the compactness of the lens. The observation of higher order relativistic images would give better bounds.

We propose that the variations of compactness against $M/D_d$ and ${\cal{D}}$ should be studied  for the following cases  as these studies would enhance our understanding of SMDOs.  First, orphaned (retro-) images of Schwarzschild mirroring (also called retro-lensing). Second, the relativistic as well as orphaned images of Kerr lensing.  The rotation parameter of the Kerr metric will decrease the value of the  compactness parameter.  The extension of our work for the Kerr metric could be also useful for a more accurate determination of rotation parameters. 
Third, naked singularities are  though abhorrent to many physicists, these are considered awesome siblings of black holes (see \cite{Vir96,VJJ97,Vir99} and references therein.) We studied gravitational lensing due to Janis-Newman-Winicour (JNW) naked singularities and obtained fascinating results \cite{VNC98,VE02,VK08}. It is important to study compactness of SMDOs when these are modeled as JNW naked singularities lenses.   The compactness parameter obtained by  gravitational lensing due to Kerr and JNW naked  singularities  spacetimes will be smaller compared to one we obtained for the Schwarzschild metric and therefore the upper bound to the compactness obtained in this paper would remain true. These extension of our work in this paper is likely to have immense implications for ngEHT project.

It is important to clarify that the observation of relativistic images of high order would very strongly support the {\em photohole}   interpretation of  massive or supermassive compact dark objects; however, we are still far behind  having any  theoretical work suggesting observations that could yield an iron-clad or even very strong  evidence of black holes. Weakly naked singularities (those covered inside at least one photon sphere \cite{VE02,VK08}) are excellent mimickers of black holes of the same ADM mass, as both  can have incredibly close sizes of their silhouettes. Therefore, silhouettes and   light rings in their vicinity would very strongly support those as photoholes, but we cannot  be sure if those are black holes. Anyway, black holes as well as weakly naked singularities  map the sources of the entire universe in the vicinity of their photon spheres.

\vspace*{-0.1in}
\acknowledgments
\vspace*{-0.2in}
This paper is  dedicated to the memory of my {\em guru}  Professor P.~C.~Vaidya. (In Sanskrit, {\em gu} means darkness and {\em ru} means one who dispels.)
Thanks are due to   F.~Eisenhauer  and H.~M.~Antia for helpful correspondence, and to G.~ F.~ R.~ Ellis for  carefully reading the manuscript. I also thank the anonymous referees for carefully reading the manuscript and for helpful suggestions.

\vspace*{0.2in}
The author has no competing interests to declare that are relevant to the content of this article.
{}
\end{document}